\documentclass[aps,pre,a4paper,reprint,showpacs,superscriptaddress]{revtex4-1}

\usepackage{amssymb}
\usepackage{color}
\usepackage{graphicx}
\usepackage{amsmath}
\usepackage{natbib}
\usepackage{pbox}

\newcommand{\reff}[1]{(\ref{#1})} 
\newcommand{\be}{\begin{equation}}
\newcommand{\ee}{\end{equation}}

\begin{document}
\begin{abstract}
Protein translation is one of the most important processes in cell life but, despite being well understood biochemically, the implications of its intrinsic stochastic nature have not been fully elucidated.
In this paper we develop a microscopic and stochastic model which describes a crucial step in protein translation, namely the binding of the tRNA to the ribosome.
Our model explicitly takes into consideration tRNA recharging dynamics, spatial inhomogeneity and stochastic fluctuations in the number of charged tRNAs around the ribosome. By analyzing this non-equilibrium system we are able to derive the statistical distribution of the times needed by the tRNAs to bind to the ribosome, and to show that it deviates from an exponential due to the coupling between the fluctuations of charged and uncharged populations of tRNA.
\end{abstract}


\title{Non-equilibrium, stochastic model for tRNA binding time statistics}
\author{Luca Caniparoli}
\affiliation{International School for Advanced Studies (SISSA), via Bonomea 265, I-34136 Trieste, Italy}
\author{Pierangelo Lombardo}
\affiliation{International School for Advanced Studies (SISSA), via Bonomea 265, I-34136 Trieste, Italy}
\affiliation{Istituto Nazionale di Fisica Nucleare, Sezione di Trieste, Italy}

\maketitle

\section{Introduction}

Recent advances in experimental physical biology are offering an unprecedented detail in the observation of the reactions happening in living systems. In fact, single molecule sensibility techniques \cite{Ritort2006,Sotomayor2007,Hummer2001,Gupta2011,Harris2007,Li2011} are beginning to probe and unveil the intrinsic stochastic nature of microscopic life.
Most notably, recent \emph{in vitro} experiments \cite{Uemura2010,Wen2008} focused on the fundamental aspects of protein translation.

Protein synthesis is one of the most common biochemical reactions happening in the cell: the individual triplets of nucleotides (the codons) composing a messenger RNA (mRNA) are translated into amino acids by the ribosomes \cite{alberts}. 
This process is biologically and chemically well understood, but the implications of its intrinsic stochastic nature have not been fully elucidated yet. 

An intriguing question concerns the ribosome dwell time distribution (DTD), i.e., the distribution of the time intervals between subsequent codon translation events.
The shape of this distribution and its dependency upon the codons heavily influence the ribosome traffic along the mRNA sequences  \cite{Mitarai2008,Reuveni2011,Greulich2012,Ciandrini2013}, and affect the efficiency, accuracy and regulation of translation \cite{Plotkin2011,Gingold2011}, as well as the process of cotranslational folding of the nascent protein \cite{OBrien2012}. 
The translation of a codon involves several subsequent biochemical steps \cite{Wen2008,Tinoco2009,Frank2010,Sharma2011}, and the stochastic duration of each of these sub-steps is typically modeled with an exponential distribution characterized by the time scale (i.e., by the rate) of that reaction \cite{Tinoco2009,Sharma2011}.
However, one among them (the \emph{binding step}) requires that the ribosome binds to an additional molecular species, the transcript RNA (tRNA), which has an internal stochastic dynamics. 

The tRNA molecules carry the corresponding amino acid to the ribosome, and physically recognize the codons effectively decoding the genetic code.
After translation has occurred and the tRNA molecule has left the ribosome, it must be recharged with the correct amino acid \footnote{The charged tRNA is a ternary complex, composed by the aminoacylated tRNA, a species-specific elongation factor, and an energy-carrying molecule (guanosine triphosphate - GTP)} before it can be used again. 
The concurrency between these two mechanisms, consumption and recharge, determines the global fraction $X$ of charged tRNA in the cell. The value of $X$ is not constant during the life of the cell, and experimental evidence showed that it can significantly vary between conditions and over time in a range from less than 1\% up to almost 100\% \cite{Dittmar2005}. It was also shown numerically that this fact can have deep consequences on translation \cite{Brackley2011,Wohlgemuth2013}.
Furthermore, the tRNA molecules have low concentrations in the cell \cite{Dong1996}: in this regime the number of tRNAs in the neighborhood of the ribosome is small, and the fluctuations in their number are relevant. 
The stochastic duration of the binding step (\emph{binding time}) is directly influenced by these fluctuations, as it depends on the concentration of charged tRNA in the neighborhood of the ribosome \cite{Uemura2010,Zhang2010}.

In order to understand how and under which conditions tRNA charging dynamics can affect the binding time distribution (BTD) (i.e., the distribution of the waiting times of the ribosome for the charged tRNA), and consequently the DTD, we develop here a stochastic model which explicitly incorporates (\emph{i}) tRNA charging and discharging dynamics, and (\emph{ii}) spatial inhomogeneity and stochastic fluctuations in the number of charged tRNAs around the ribosome. 
This minimal model captures these two fundamental aspects of the translation process \footnote{We did not consider for instance the enzymatic nature of the recharging of the tRNAs, the continuous spatial dependency of the tRNA density, nor tRNA proofreading.}, and is analytically tractable.
Its solution, validated using Monte Carlo numerical simulations, shows that the interplay between diffusion, recharging and translation dynamics induces a coupling between the fluctuations in the number of charged and uncharged tRNAs. 
Due to this phenomenon the BTD, which we obtain analytically from the model, deviates from a pure exponential, consistently with the findings in Ref.~\cite{Zhang2010}.
Besides, this model asymptotically reaches a non equilibrium steady state (NESS). NESSs have attracted a lot of attention since a variety of systems in physics, chemistry, biology and engineering exhibit them, and their characterization is typically far more difficult than the equilibrium states \cite{Zia2006,Zia2007,Platini2011,Chou2011}.

The structure of the paper is as follows: after defining the model in Sec.~\ref{model}, we characterize the stationary state in Sec.~\ref{sec:stat}. The BDT is obtained in Sec.~\ref{sec:BTD}, where its main features are analyzed. In the last subsection we show how an additional biochemical step can be included, in order to get an estimate for the DTD. In Sec.~\ref{parameters}, we discuss the interpretation of the parameters of the model in terms of measurable quantities and we produce, when possible, an order-of-magnitude estimate for their values.
We conclude by reviewing and commenting our results in Sec.~\ref{conclusions}.

\section{The model}
\label{model}

We model a ribosome translating an mRNA (a string of codons) into a protein (a string of amino acids), with the scope of analyzing the effects of tRNA charging dynamics and its finite availability on translation dynamics. 

The fraction of charged tRNAs in the cell has a very wide variation range (up to 2 orders of magnitude, depending on the tRNA species and growth conditions \cite{Dittmar2005}) and exclusively affects the binding step.
For this reason we focus here on this specific step, neglecting all the other biochemical reactions, which will be accounted for in Sec.~\ref{other_steps_section}.
For simplicity, each translation event is assumed to be instantaneous: as the charged tRNA  binds to the ribosome, (i) it is uncharged and released in the system, (ii) the codon is translated and the ribosome translocates to the next codon.
Besides, we treat the special case of a single tRNA species translating a single type of codons.

The ribosome consumes charged tRNAs during translation. On average, the concentration of charged tRNAs is lower close to the ribosome and rises increasing the distance, as shown in Ref.~\cite{Zhang2010}. 
In order to model this spatial inhomogeneity, we suppose that the ribosome can recruit the tRNAs within an effective distance $r$, i.e., in an effective volume $V_r=4\pi r^3/3$.
The tRNAs farther than $r$ are considered as part of an infinite reservoir, and can be exchanged with the system due to diffusion. The concentrations of charged tRNAs within the volume $V_r$ is different from that in the reservoir, and it is determined by the stochastic translation dynamics. 

\begin{figure}
	 \centering\includegraphics[width=\columnwidth]{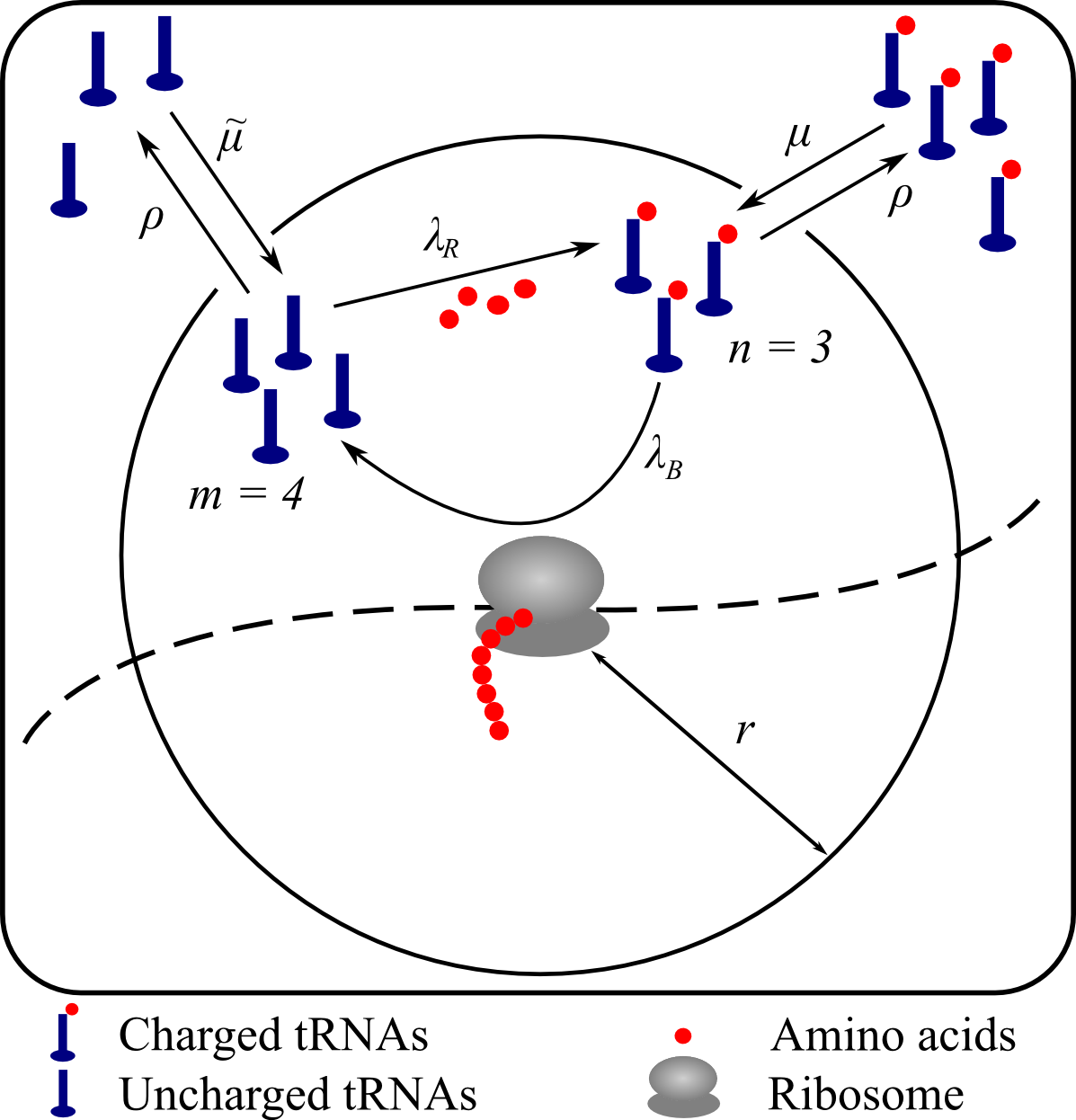}
	 \caption{(Color online) Cartoon of the model representing the possible reaction pathways. Uncharged tRNAs (blue, on the left) can either be exchanged with the reservoir or be recharged (with rate $\lambda_R$), illustrated by the addition of an amino acid (red dots). Similarly, the charged tRNAs can enter in or leave the system, or bind (with rate $\lambda_B$) to the ribosome (gray), which translates an mRNA (dashed line). In the state represented here, $n=3$ and $m=4$.}
	 \label{system}
\end{figure}

Let us therefore consider a system which comprises a ribosome (translating an mRNA composed by several repeats of the same codon), $n$ charged and $m$ uncharged tRNAs (see Fig.~\ref{system}).
Each uncharged tRNA can be recharged with rate $\lambda_R$, while each charged tRNA can bind to the ribosome with rate $\lambda_B$, becoming uncharged.
We also suppose that there is a stochastic flux between the system and the infinite reservoir, i.e., that each tRNA can exit the system with rate $\rho$, while with rate $\mu$ ($\tilde\mu$) a charged (uncharged) tRNA diffuses from the reservoir into the system. 
We refer to $\rho$ as the \emph{diffusion rate}, since the exit rate from the volume $V_r$ is determined by how fast the Brownian diffusion is (as we discuss in Sec.~\ref{parameters}).
This subdivision in system-reservoir encodes, in the most simple way, the spatial inhomogeneity of the charged tRNA fraction close to the ribosome.

%
%

Considering an infinitesimal time interval $\delta t$, the possible transitions (with the corresponding probabilities) are:
\begin{itemize}
	\item $(n,m)\overset{m \lambda_R\,\delta t}{\longrightarrow}(n+1,m-1)$: recharge,  one uncharged tRNA  gets charged.
	\item $(n,m)\overset{n \lambda_B\,\delta t}{\longrightarrow}(n-1,m+1)$: binding, one charged tRNA gets discharged and one codon is translated.
	\item $(n,m)\overset{\mu\,\delta t}{\longrightarrow}(n+1,m)$ and $(n,m)\overset{\tilde\mu\,\delta t}{\longrightarrow}(n,m+1)$: a tRNA (respectively charged, uncharged) enters the system from the reservoir.
	\item $(n,m)\overset{n\rho\,\delta t}{\longrightarrow}(n-1,m)$ and $(n,m)\overset{m\rho\,\delta t}{\longrightarrow}(n,m-1)$: a tRNA (respectively charged, uncharged) leaves the system.
\end{itemize}
These rates define, in general, a non-equilibrium system: the stationary state is a function of all the rates, as we show in the next section.

\section{Stationary distribution of the number of charged tRNAs}
\label{sec:stat}
The set of rates given in the previous section produces the following master equation for the probability $p_{n,m}(t)$ of being in the state $(n,m)$:
\be\label{master_eq_1codon}
\begin{split}
 \dot p_{n,m}=&\ \lambda_R[-m\,p_{n,m}+(m+1)p_{n-1,m+1}]\\
 &+\lambda_B[-n\,p_{n,m}+(n+1)p_{n+1,m-1}]\\
&-(\mu+\tilde\mu)p_{n,m}+\mu\,p_{n-1,m}+\tilde\mu\,p_{n,m-1}\\
&+\rho[(n+1)p_{n+1,m}+(m+1)p_{n,m+1}\\
&-(n+m)p_{n,m}].
\end{split}
\ee
We focus on the stationary state of the system by setting $\dot p_{n,m}=0$. Since the system is ergodic, the stationary state is unique and it is reached after a relaxation time that will be discussed forward in this section.

In order to determine the stationary solution of Eq.~\reff{master_eq_1codon}, we introduce the generating function $G(z,w)=\sum_{n,m=0}^{\infty}p_{n,m}z^n w^m$ and we obtain
\begin{multline}
\label{gen_function_1codon}
	  \lambda_R(z-w)\partial_w G+\lambda_B(w-z)\partial_z G+\rho[(1-z)\partial_zG\\
  +(1-w)\partial_w G]+\mu(z-1)G+\tilde\mu(w-1)G=0,
\end{multline}
whose solution can be calculated by using the method of the characteristics. After imposing the condition $G(1,1)=1$ (normalization), we have
\be
 G(z,w)=e^{\frac{(z-1)[\lambda_R(\mu+\tilde\mu)+\mu\rho]+(w-1)[\lambda_B(\mu+\tilde\mu)+\tilde\mu\rho]}{\rho(\lambda_R+\lambda_B+\rho)}},
\ee
and by recursive differentiation, we obtain the stationary probability
\be\label{pnm}
 p_{n,m}= \left[\frac{(\partial_z)^n(\partial_w)^m}{n!\,m!}G(z,w)\right]_{\substack{z=0\\w=0}}=
 \frac{e^{-\bar N}\,\bar n^n\,\bar m^m}{n!\,m!},
\ee
where $\bar n,\ \bar m$ and $\bar N$ are the average values of the quantities $n$, $m$ and $N=n+m$, respectively:
\be
 \begin{split}\label{valori_medi}
   &\bar n=\langle n\rangle=\frac{\lambda_R\,\bar N+\mu}{\lambda_B+\lambda_R+\rho},\\
   &\bar m=\langle m\rangle=\frac{\lambda_B\,\bar N+\tilde\mu}{\lambda_B+\lambda_R+\rho},\\
   &\bar N=\langle n+m\rangle=\frac{\mu+\tilde\mu}{\rho}.
 \end{split}
\ee
The stationary distribution Eq.~\reff{pnm} is a factorized Poissonian in $n$ and $m$ \footnote{Note that, since detailed balance does not hold in general, the system is out of equilibrium and it is not a priory obvious to find a stationary distribution \cite{Zia2007,Platini2011}}: the two variables are uncorrelated \emph{at the same time} (we anticipate that the same is not true for \emph{different times}, as  we show in Sec.~\ref{sec:BTD}).

Using the last of Eq.s~\reff{valori_medi}, the parameters $\mu$ and $\tilde\mu$ can be conveniently expressed in terms of the diffusion parameter $\rho$ and of the average tRNA number $\bar N$:
\begin{align*}
 \mu&=X \bar N\rho, \\
 \tilde\mu&=(1-X )\bar N\rho,
\end{align*}
where we introduced the parameter $0\leq X\leq 1$ which measures the fraction of charged tRNA in the reservoir. Note well that $X$ was measured \emph{in vivo} in Ref.~\cite{Dittmar2005}.

In order to simplify the notation, let us rescale the time such that $\lambda_B=1$, and set $\lambda_R=\lambda$.
Let us also introduce the average fraction $x$ of charged tRNAs into the system:
\be\label{x}
x\equiv\frac{\lambda\,+X \rho}{1+\lambda+\rho}.
\ee
The average values for $n$ and $m$ can be expressed as $\bar n=\bar N x$ and $\bar m=\bar N (1-x)$.

These quantities and the stationary distribution Eq.~\reff{pnm} behave as expected in the limit $\rho\to\infty$: the system is at equilibrium with the cell and the average fraction $x$ of charged tRNA therein coincides with the fraction $X$ in the cell: $x=X$.
On the  other hand, if $\rho\to 0$, the diffusion is much slower than binding, and the average number of charged tRNAs is completely determined by the internal dynamics: $x=\lambda (1+\lambda)^{-1}$. In this case the effect of diffusion amounts to a slow but not negligible fluctuation of the tRNA number $N=n+m$.

The exponential relaxation to the stationary distribution is ruled by the two time scales $t_1=\rho^{-1}$ and $t_2=(\lambda+\rho+1)^{-1}$, which are deduced from the time-dependent solution of Eq.~\reff{master_eq_1codon} (the solution is given in App.~\ref{app:time_dep}). The stationary state is reached when the observation time is larger than the largest of these time scale: $T_{\rm obs}\gg \rho^{-1}$.

Finally, we observe that the detailed balance condition is satisfied only for $\rho=0$ (i.e., in the absence of diffusion), or for $\lambda=X/(1-X)$ (see App.~\ref{app:DB} for the proof). In the latter case the stationary average values for the charged fraction of tRNA of both the internal and the diffusive dynamics coincide, and $x=X$. Apart from these two special points, the stationary state is a non-equilibrium state.

\section{Statistics of binding times}
\label{sec:BTD}

The average binding time per codon predicted by this model is trivially $1/\bar n$. In general, however, when the distribution is not exponential, the average does not fully characterize the behavior of the random variable. In this section we therefore compute analytically the probability density function for the intervals between two subsequent binding events.

The derivation is carried out by writing a master equation which accounts for an auxiliary variable $s$ counting the number of time steps elapsed since the last binding event (see below). This procedure allows the calculation of the cumulative distribution of the binding times and finally of the BTD.

Let us consider a discrete-time dynamics where $\delta t$ is the unit time interval. The state of the system is described by $(n,m;s)$, where the counter $s$, at each time interval, is either set to zero if a binding event occurs, or increased by one otherwise. 
Without loss of generality, we set $\lambda_B=1$ from the beginning. The possible transitions are:
\begin{itemize}
	\item $(n,m;s)\overset{m \lambda\,\delta t}{\longrightarrow}(n+1,m-1;s+1)$: one uncharged tRNA  gets recharged
	\item $(n,m;s)\overset{n \,\delta t}{\longrightarrow}(n-1,m+1;0)$: one codon is translated and one charged tRNA gets discharged
	\item $(n,m;s)\overset{\mu\,\delta t}{\longrightarrow}(n+1,m;s+1)$ and $(n,m;s)\overset{\tilde\mu\,\delta t}{\longrightarrow}(n,m+1;s+1)$: a tRNA (respectively charged, uncharged) enters the system from the reservoir
	\item $(n,m;s)\overset{n\rho\,\delta t}{\longrightarrow}(n-1,m;s+1)$ and $(n,m;s)\overset{m\rho\,\delta t}{\longrightarrow}(n,m-1;s+1)$: a tRNA (respectively charged, uncharged) leaves the system to the reservoir
	\item $(n,m;s)\overset{1-\delta t[n+\lambda  m+\mu+\tilde\mu+\rho n+\rho m]}{\longrightarrow}(n,m;s+1)$: nothing happens and the counter is increased.
\end{itemize}
\begin{widetext}
This set of rates leads to the discrete time master equation for the probability $q_{n,m;s}(t)$ of being in the state $(n,m;s)$ at time $t$
 \begin{multline}
   \frac{q_{n,m;s}(t+\delta t)}{\delta t}= \lambda (m+1)\,q_{n-1,m+1;s-1}(t)+\mu\, q_{n-1,m;s-1}(t)+\tilde\mu\,q_{n,m-1;s-1}+\rho(n+1)q_{n+1,m;s-1}(t)\\
   +\rho(m+1)q_{n,m+1;s-1}+\frac{q_{n,m;s-1}(t)}{\delta t}
   -\big[ n+\lambda m+\mu+\tilde\mu+\rho n+\rho m\big]q_{n,m;s-1}(t)+\delta_{s,0}\sum_{s'=0}^{\infty}(n+1)q_{n+1,m-1;s'}(t).
 \end{multline}
The limit $\delta t\to 0$ is well defined by setting $\tau=s\delta t$ and it results in the following partial differential equation:
\begin{multline}
	\partial_tq_{n,m}(\tau,t)= -\partial_{\tau}q_{n,m}(\tau,t)+\lambda (m+1)q_{n-1,m+1}(\tau,t)-( n+\lambda m)q_{n,m}(\tau,t)
+\rho\big[(n+1)q_{n+1,m}(\tau,t)\\
+(m+1)q_{n,m+1}(\tau,t)-(n+m)q_{n,m}(\tau,t)\big]+\mu\, q_{n-1,m}(\tau,t)+\tilde\mu\,q_{n,m-1}(\tau,t)-(\mu+\tilde\mu)q_{n,m}(\tau,t)+(n+1)\delta(\tau)p_{n+1,m-1}(t),
\end{multline}
where $p_{n,m}(t)=\int_0^{\infty}d\tau\, q_{n,m}(\tau,t)$ is the solution of Eq.~(\ref{master_eq_1codon}).
\end{widetext}

The differential equation for the stationary probability is obtained by setting $\partial_t q_{n,m}(\tau,t)=0$, and reads
\begin{multline}
	\label{ME_time-distribution_stationary}
  \partial_{\tau}q_{n,m}(\tau)= \lambda (m+1)q_{n-1,m+1}(\tau)\\
  -( n+\lambda m)\,q_{n,m}(\tau)+\rho\Big[(n+1)q_{n+1,m}(\tau)+(m+1)q_{n,m+1}(\tau)\\+X  \bar N\, q_{n-1,m}(\tau)
  +(1-X ) \bar N\, q_{n,m-1}(\tau)\\
  -(n+m+ \bar N)q_{n,m}(\tau)\Big]+\delta(\tau)\,\alpha_{n+1,m-1},
\end{multline}
where $\alpha_{n,m}=\,n\,p_{n,m}$ and $p_{n,m}$ is provided by Eq.~\reff{pnm}.

Similarly to the previous case, we introduce the generating function
\be
 G(z,w;\tau)=\sum_{n=0}^{\infty}\sum_{m=0}^{\infty}q_{n,m}(\tau)z^n w^m,
\ee
and Eq.~\reff{ME_time-distribution_stationary} becomes
\begin{multline}
\label{stationary_gen-function_solution}
 \partial_{\tau}G=\lambda (z-y)\partial_y G- z\partial_z G\\
 +\rho\Big[(1-z)\partial_z+(1-w)\partial_w+\bar N X (z-1)\\
 +\bar N (1-X) (w-1)\Big]G+\delta(\tau)f(z,w),
\end{multline}
where
\begin{equation}
\begin{split}
  f(z,w)=&\sum_{n=0}^{\infty}\sum_{m=1}^{\infty}\alpha_{n+1,m-1}z^n w^m\\
  =&\, \bar n\,w\,\exp[\bar n(z-1)+\bar m(w-1)].		
\end{split}
\end{equation}
Even though Eq.~\reff{stationary_gen-function_solution} could be solved in full generality, here we are interested in the particular value $G(1,1,\tau)$ as it coincides with the marginal distribution
\be
Q(\tau)=\sum_{n,m=0}^{\infty}p_{n,m}(\tau)=G(1,1;\tau)
\ee
for $\tau$. 
The probability $P(\tau)$ for the time interval $t$ between two subsequent binding events to be $t>\tau$, is proportional to $Q(\tau)$. 
In fact, let us suppose that, at some time $T$ during the evolution of the system, the auxiliary variable has a value $\tau=\tau^*$. In this case, the time interval $t$ between the two subsequent binding events enclosing $T$ is, by construction, $t>\tau^*$. It follows that $P(\tau)=Q(\tau)/Q(0)$.

By solving Eq.~\reff{stationary_gen-function_solution} with $w=z$, we obtain the generating function $G(z,z;\tau)$ and the probability $P(\tau)$, describing the probability for the time $t$ between two consecutive binding events to be larger than $\tau$:
\begin{multline}
\label{P(tau)_1codon}
  P(\tau)= \left[ R+ \frac{\lambda}{\lambda-1}\left(\frac{e^{-\tau(\rho+1)}}{\rho+1}-\frac{e^{-\tau(\rho+\lambda)}}{\rho+\lambda}\right)\right] e^{-\tau R \bar N x}\\
\times\exp\Big[\frac{\lambda\bar N x}{\lambda-1}
   \left(\frac{e^{-\tau(\rho+1)}-1}{(\rho+1)^2}-\frac{e^{-\tau(\rho+\lambda)}-1}{(\rho+\lambda)^2}\right) \Big],
\end{multline}
where
\be
R=\frac{\rho(\rho+\lambda+1)}{(\rho+1)(\rho+\lambda)}
\ee
is always $<1$, and $x$ is the fraction of charged tRNAs in the system, Eq.~\reff{x}.

For further reference, note that the function $P(\tau)$ can be written as $P(\tau)=\partial_\tau A(\tau)$, with
 \begin{multline}
 \label{A_t}
 A(\tau)=-\frac{1}{\bar N x}\exp\Big[-\tau R\bar N x\\
 +\frac{\lambda\bar N x}{\lambda-1}\left(\frac{e^{-\tau(\rho+1)}-1}{(\rho+1)^2}-\frac{e^{-\tau(\rho+\lambda)}-1}{(\rho+\lambda)^2}\right) \Big].
\end{multline}

\begin{figure}
	 \includegraphics[width=\columnwidth]{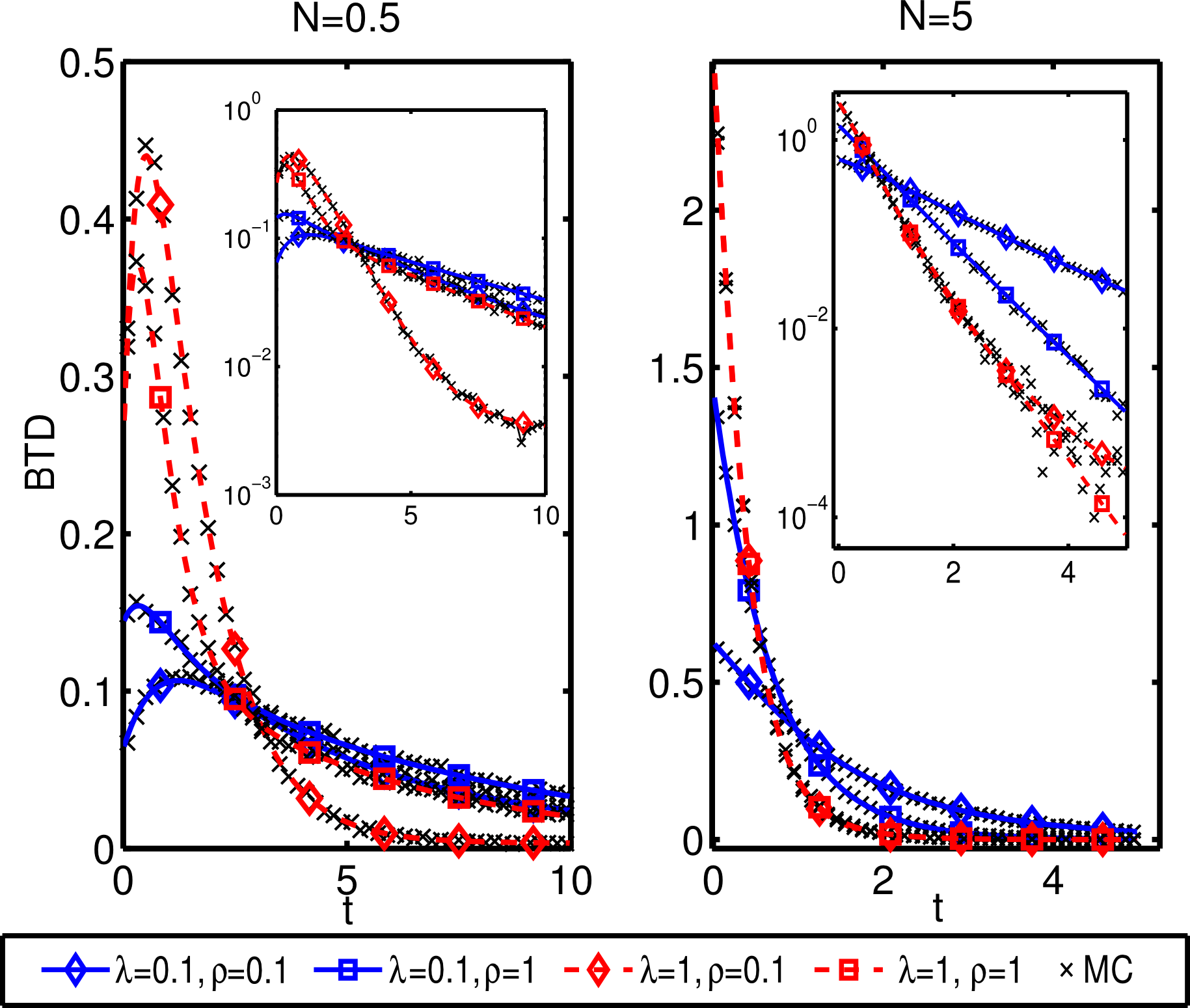}
	 \caption{(Color online) Probability density function $p(t)$ for the binding time (BTD) for various choices of the parameters $\lambda$ (blue and continuous lines for $\lambda=0.1$, red and dashed lines for $\lambda=1$), $\rho$ (diamonds for $\rho=0.1$, squares for $\rho=1$) and $\bar N$ ($\bar N=0.5$ in the left panel, $\bar N=5$ in the right one). The parameter $X$ is kept fixed to $0.5$. The black crosses are the results of Monte Carlo simulations, and do not show any significant deviation from the theoretical predictions. As the two log-plot insets show, deviations from a pure exponential are most evident for small $\bar N$, where the fluctuations play a relevant role.
}
	 \label{BTD}
\end{figure}

Let us now observe that $P(\tau)$ is the complement of the cumulative distribution for the BTD, defined as $p(t)$. Therefore, since $Q(\tau)=\int_\tau^\infty dt\,p(t)$, the BTD is given by
\be
\label{p_t}
 p(t)=-\partial_\tau P(\tau)_{|\tau=t}.
\ee
Some typical realizations of  $p(t)$ are shown in Fig.~\ref{BTD}, where we also compare the theoretical prediction with the numerical Monte Carlo simulations. We did not observe any significant deviation between the theoretical results  and the simulations. Interestingly, for small times and small values of $\bar N$ the BTD relevantly deviates from an exponential (see the log-plot insets of Fig.~\ref{BTD}). On the other hand, these deviations are milder for small values of $\lambda$ and large values of $\rho$. The main features the $p(t)$ are analyzed in the next section.

\subsection{Characterization of the BTD}

In order to characterize the BTD $p(t)$, we calculate its first two moments. We compare the second moment of the BTD with that from an exponential distribution having the same mean, observing that the BTD is overdispersed with respect to that distribution.  

The first moment -i.e., the average of the BTD- is given by
\be
\label{avg}
 \langle t\rangle=\int_0^{\infty}dt\,t\,p(t)=\int_0^{\infty}d\tau\,P(\tau)=\frac{1}{\bar N x},
\ee
and coincides with the inverse of the average number $\bar n$ of charged tRNAs in the system, as expected.

The second moment is given by  
\begin{equation}
	\langle t^2\rangle= \int_0^{\infty}dt\,t^2\,p(t)=2\int_0^{\infty}dt\,t\,P(t)=-2\int_0^{\infty}dt\,A(t),
\end{equation}
where $A(t)$ is given by Eq.~\reff{A_t}, and can be written as
\begin{multline}
\label{t2}
  \langle t^2\rangle=\frac{2}{\bar N x}\exp\left[\frac{\lambda\bar N x}{\lambda-1}\left(\frac{1}{(\rho+\lambda)^2}-\frac{1}{(\rho+1)^2}\right)\right]\\
  \times\int_0^1dy\,y^{R\bar N x-1}\exp\left[\frac{\lambda\bar N x}{\lambda-1}\left(\frac{y^{\rho+1}}{(\rho+1)^2}-\frac{y^{\rho+\lambda}}{(\rho+\lambda)^2}\right)\right].
 \end{multline}
Equation \reff{t2} can be numerically evaluated in order to determine the variance $\sigma_t^2=\langle t^2\rangle-\langle t\rangle^2$.

\begin{figure}
	\includegraphics[width=\columnwidth]{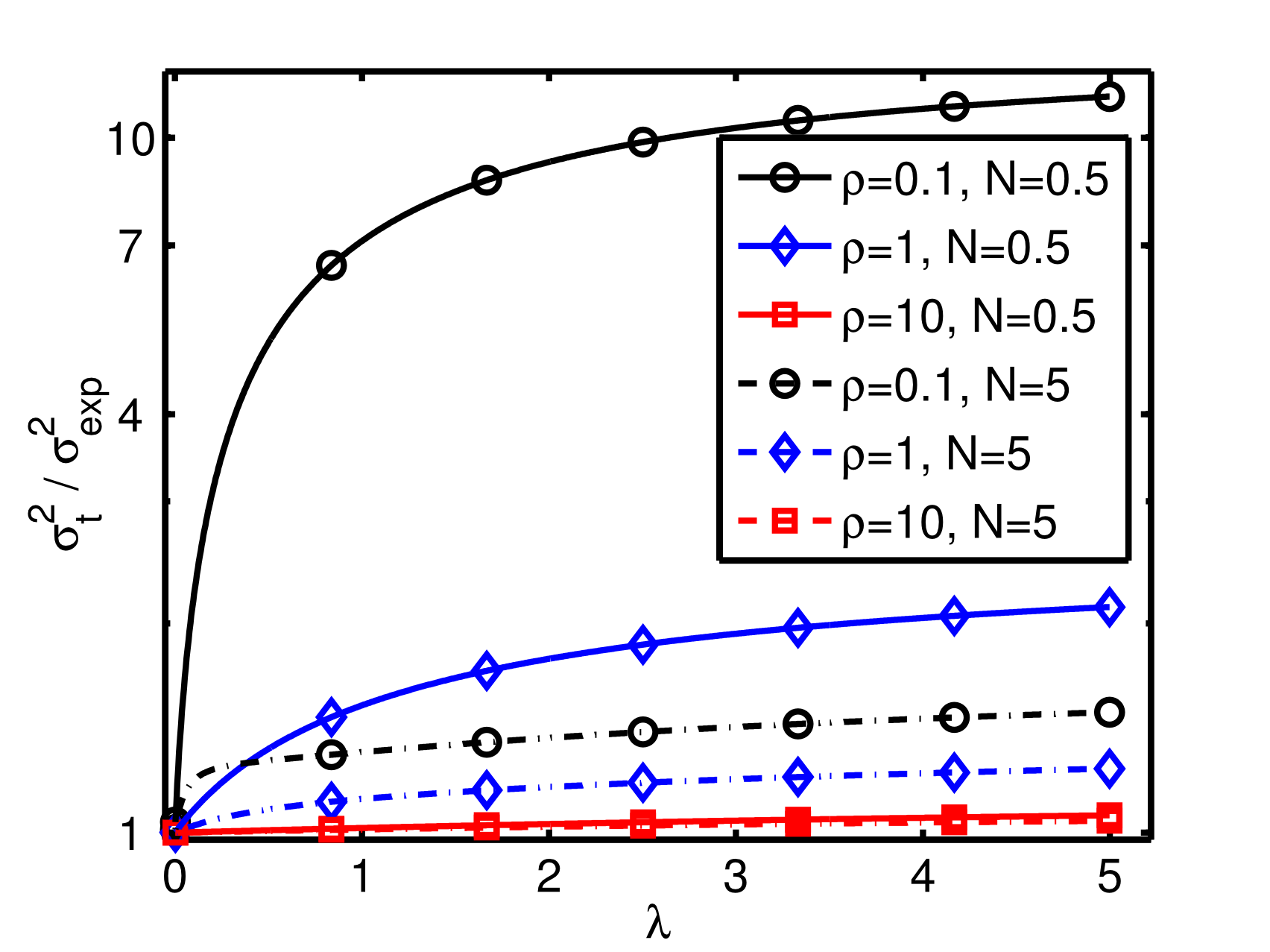}
	 \caption{(Color online) Ratio between the variances $\sigma_t^2$ of the BTD, and $\sigma_{\rm{exp}}^2$ of the exponential with the same average Eq.~\reff{pexp}, as a function of $\lambda$, for various values of $\bar N$ and $\rho$ (the parameter $X$ do not change qualitatively the results and is set to $X=1/2$). For $\lambda\to 1$ the BTD converges to an exponential distribution, as shown by Eq.~\reff{smalltimesexp}, and the ratio $\sigma_t^2/\sigma_{\rm{exp}}^2\to 1$. In general, however, the BTD is over dispersed with respect to the exponential distribution.}
	 \label{fig:var}
\end{figure}

In Fig.~\ref{fig:var} we plot the ratio $\sigma_t^2/\sigma_{\textrm{exp}}^2$ for various values of the parameters, where $\sigma_{\textrm{exp}}$ is the variance of the exponential distribution
\be\label{pexp}
 p_{\rm exp}(t)=x\bar N e^{-x\bar N t},
\ee
fixed to having the same average of the BTD. By inspection, we did not find any point in the parameter space such that $\sigma_t<\sigma_{\textrm{exp}}$: the BTD is overdispersed with respect to the exponential distribution, Eq.~\reff{pexp}. 

This observation can be further characterized by comparing the small and large $t$ expansions of the two distributions: first, by analyzing the Taylor expansion around $t=0$ of the two probability distributions, we observe that $p(t)-p_{\textrm{exp}}(t)\sim \lambda t + O(t^2)$.
Short binding times are under represented in the exponential distribution. Also note that for $\lambda=0$ the two distributions coincide and the ratio $\sigma_t/\sigma_{\textrm{exp}}=1$, as shown in Fig.~\ref{fig:var}.

The tails of the two distributions also differ in the large $t$ limit. In the $t\to \infty$ limit, Eq.~\reff{p_t} behaves as
\be
\label{large_t}
	p(t) \propto e^{-Rx\bar Nt}.
\ee
Comparing this expression with Eq.~\reff{pexp} and noting that $R<1$ (by definition), we see that large binding times are under represented in the exponential distribution.

Finally, the BTD in Eq.~\reff{p_t} reduces to an exponential in the slow recharge limit $\lambda\to 0$, where
\be
\label{smalltimesexp}
 p(t)\to x\bar N e^{-x\bar N t},
\ee
and in the fast diffusion limit $\rho\to\infty$:
\be
 p(t)\to X  \bar N e^{-X  \bar N t}.
\ee
In the latter case the charged fraction $x$ of tRNA in the system coincides with the fraction $X$ in the reservoir, consistently with the expectation that in the fast diffusion limit the fluctuations of charged tRNAs are determined by the exchange with the bath and are uncorrelated in time.
As we show in the next section, these two limits have an interesting physical interpretation.

\subsection{The BTD deviates from an exponential due to the time-correlations of $n$ and $m$}

The time evolution of the model introduced in Sec.~\ref{model}, by exclusively depending on the present state, is Markovian and memoryless. 
Markovianity typically implies exponentially distributed time intervals between events (as the exponential is the only memoryless distribution \cite{feller}), and the deviation of the BTD from the exponential in this model could be surprising at first sight. 
Here we show that this deviation arises due to a nontrivial coupling between the fluctuations of $n$ and $m$. 

First, let us introduce the average value  $\langle n(t)\rangle_{\rm nb}$ of $n$ at time $t$ after a binding event (at time $t=0$), conditioned to the fact that no other binding events were recorded up to time $t$.
The BTD is related to $\langle n(t)\rangle_{\rm nb}$ by
\be\label{BTD_nt}
 p(t)= \langle n(t)\rangle_{\rm nb}\exp\left(-\int_0^t dt'\, \langle n(t')\rangle_{\rm nb}\right),
\ee
as proved in App.~\ref{app:BTD_n}. Interestingly, Eq.~\reff{BTD_nt} shows that the deviations from an exponential of the BTD appear as soon as $\langle n(t)\rangle_{\rm nb}$ departs from a constant and acquires a time dependency.

\begin{figure}
	 \includegraphics[width=\columnwidth]{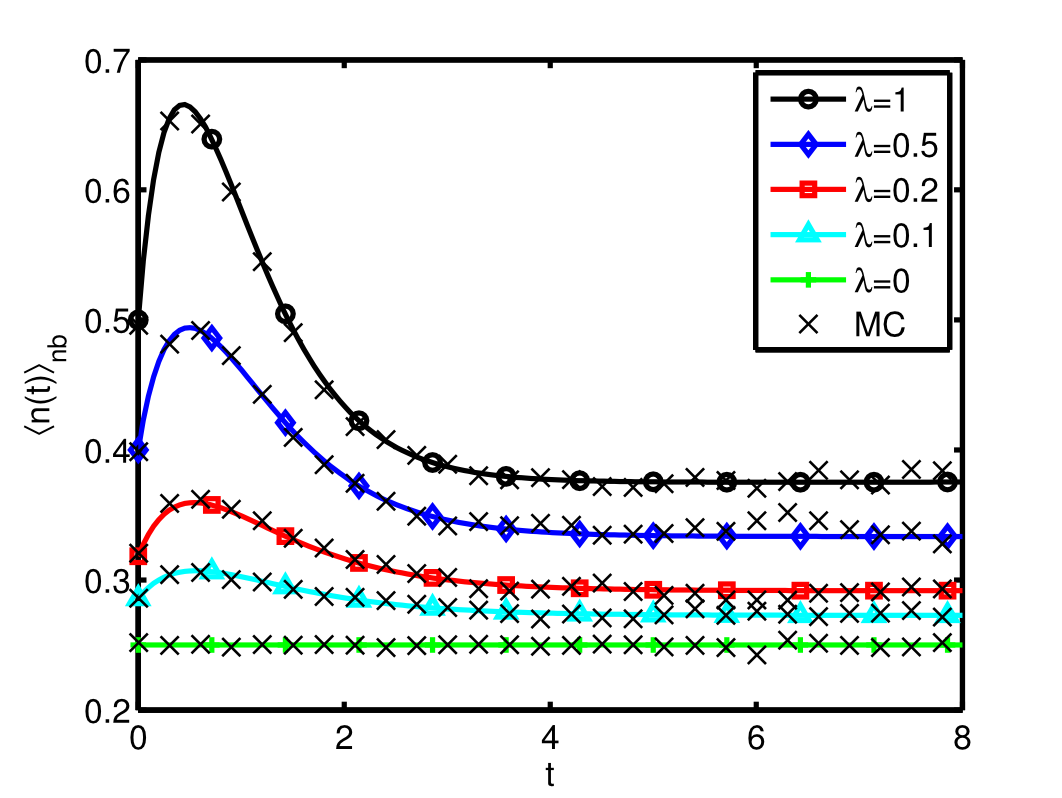}
	 \caption{(Color online) Average number of charged tRNAs in the system $\langle n(t)\rangle_{\rm nb}$ conditioned to the fact that no binding event occurred up to time $t$ after that one at $t=0$, plotted for various values of $\lambda$, evaluated analytically (as from Eq.~\reff{nAvgCond_P}, solid lines), and numerically by Monte Carlo (black crosses). The other parameters were set to $\bar N =1$, $\rho=1$ and $X=1/2$.  As $\lambda$ approaches 0, the function $\langle n(t)\rangle_{\rm nb}$ becomes a constant.}
	 \label{n_avg_conditioned}
\end{figure}

In the stationary regime, a binding event occurs with probability proportional to $np_n$, where $p_n=\sum_m p_{mn}$ is the marginal stationary probability for $n$, obtained from Eq.~\reff{pnm}. Precisely, the distribution for $n$ at the instant before a binding event is:
\be
p_n^{\textrm{b}^-}=\frac{\bar n^{n-1}}{(n-1)!}e^{-\bar n},
\ee
whose average is $\bar n_{{\rm b}^-}=\bar n+1$: a binding event typically happens when a fluctuation rises the number of charged tRNAs close to the ribosome (i.e., within the volume $V_r$). Note well that the number $m$ of uncharged tRNAs is not influenced.
After the translation event $n\to n-1$ and $m\to m+1$. Therefore, immediately after translation, $\bar n_{{\rm b}^+}=\bar n$ and $\bar m_{{\rm b}^+}=\bar m+1$:  the fluctuation on $n$ has propagated to $m$.
Now, if $\lambda>0$ and $\rho<\infty$, this fluctuation can again propagate to $n$ with a characteristic time scale, producing a loop which induces a time dependency in $\langle n(t)\rangle_{\rm nb}$. This mechanism is suppressed if $\lambda=0$ and it is negligible if $\rho\to\infty$. In the former case, in fact, the dynamics of $n$ is not affected by the dynamics of $m$ (as it can easily be seen from the rates described at the beginning of Sec.~\ref{model}).
This intuition is confirmed by the numerical simulations in Fig~\ref{n_avg_conditioned}, where it is shown that the average $\langle n(t)\rangle_{\rm nb}$ reduces to a constant for $\lambda\to 0$. For $\rho\to\infty$ the fluctuations on $m$ are immediately dissipated in the thermal bath before they can propagate back to $n$.

As a further check, we can quantify the influence of $m$ at a given time on the future $n$ dynamics -and vice versa- by studying the two point  correlators
\begin{align}
C_{mn}(t)&\equiv\langle m(0)n(t)\rangle-\bar n\bar m=\bar m\lambda\frac{e^{-\rho t}-e^{-t(\lambda+\rho+1)}}{\lambda+1},\label{corr_mn}\\
  C_{nm}(t)&\equiv\langle n(0)m(t)\rangle-\bar n\bar m=\bar n\frac{e^{-\rho t}-e^{-t(\lambda+\rho+1)}}{\lambda+1},\label{corr_nm}
\end{align}
whose derivation is carried out in App.~\ref{app:time_dep}.

The correlation between $m(\tau)$ and $n(\tau+t)$, Eq.~\reff{corr_mn}, vanishes identically when $\lambda=0$ or $\rho\to\infty$. The dynamics of $n$ decouples from that of $m$ because the fluctuations of $m$ cannot propagate to $n$. Note, however, that the reverse is not true: the two variables are not independent  when $\lambda=0$, as shown by the fact that the correlator $C_{nm}(t)$ in Eq.~\reff{corr_nm} does not vanish.
On the other hand, for $\lambda>0$ and $\rho<\infty$, the correlator Eq.~\reff{corr_mn} is a linear combination of two exponentials with different decay times. It is not monotonic in $t$, as it vanishes at $t=0$ (as expected from the factorization of the stationary probability) and has a maximum for $t=(1+\lambda)^{-1}\log[\rho^{-1}(\lambda+\rho+1)]$.

Let us finally observe that the exponential BTD limits ($\lambda=0$ and $\rho\to \infty$), are coherent with the absence of memory in the time series of $n$. The BTD is directly dependent only on the time series of $n$, and not on $m$. In the process in which only the $n$ variable is observed (i.e., the projection of the original process on the $n$ variable) the information on $m$ is missing. 
Therefore, the future evolution of $n$ is not completely determined by its present state, and its behavior is not Markovian.  In the two limits $\lambda=0$ and $\rho\to\infty$, $n$ is effectively decoupled from the evolution of $m$: the timeseries of $n$ is Markovian and memoryless, and the BTD is exponential, as the exponential is the only memoryless continuous distribution \cite{feller}.
The complete model, however, is always Markovian, as its time evolution exclusively depends on the current state $(n,m)$ and not on the past history.

\subsection{Other biochemical steps}\label{other_steps_section}

The derivation of the BDT $p(t_B)$, Eq.~\reff{p_t}, was carried out in the limit where the binding step is rate limiting, i.e., by neglecting the additional biochemical steps which are required to translate a codon \cite{Tinoco2009,Sharma2011}. Here we show how these additional steps can be considered using a simple approximation, once a time distribution describing their duration is known.

Let us split the dwell time $t_D$ of the ribosome into the binding time $t_B$, and the time $t_A$ spent for the additional biochemical reactions, such that $t_D=t_B+t_A$.
As a first approximation, let us assume that $t_A$ is described by a Poisson process with rate $\lambda_A$, modeling one additional biochemical step. The times $t_A$ are drawn from the exponential distribution 
\begin{equation}
p_A(t_A)=\lambda_A e^{-\lambda_A t_A}.
\label{exp_A}
\end{equation}
During $t_A$ the binding is suppressed, while recharge and exchange with the reservoir continue normally. Since the duration of $t_A$ influences the tRNA charging level in the binding step, the two time intervals $t_A$ and $t_B$ are not independent random variables, and we expect that the distribution $p(t_B)$ is not accurately described by Eq.~\reff{p_t} any more. 

At the mean field level, the influence of $t_A$ on Eq.~\reff{p_t} can be treated by rescaling the rates to the effective values
\be\label{effective_parameters}
\lambda^{\rm e}=\gamma\lambda,\hspace{5mm} \rho^{\rm e}=\gamma \rho,
\ee
with
\be\label{gamma}
 \gamma=\frac{\bar t_A+\bar t_B}{\bar t_B},
\ee
where $\bar t_A$ and $\bar t_B$ are the mean values of $t_A$ and $t_B$, respectively. This is equivalent to assume that recharge and diffusion occur during the binding step only, with effective rates given by Eq.~\reff{effective_parameters}. This simplified approach is most accurate if $\bar t_B\gg \bar t_A$ (i.e., when the binding step is rate limiting \cite{Liljenstroem1985}) or if $\bar t_B\gg 1/\rho$ (i.e., when diffusion is very fast and therefore, after each translation, $n(t)$ quickly reaches its stationary value).

The mean binding time $\bar t_B$ in Eq.~\reff{gamma} can be computed by plugging the effective parameters $\lambda^{\rm e}$ and $\rho^{\rm e}$ in Eq.~\reff{avg}, and by substituting this value into Eq.~\reff{gamma}. Then, solved the equation for $\gamma(\lambda^{\rm e},\rho^{\rm e})$, and substituted this value in Eq.s \reff{effective_parameters}, we obtain the bare parameters ($\rho$, $\lambda$) as functions of the rescaled ones ($\rho^{\rm e}$, $\lambda^{\rm e}$). By inverting these relations, we compute the rescaling factor $\gamma$ as a function of the  rate $\lambda_A$ and of the bare parameters:
\be
 \begin{split}
   &\gamma=\frac{ k+\sqrt{k^2+4(\lambda+\rho)}}{2(\lambda+\rho)},\\
 \end{split}
\ee
where $k=\lambda+\rho-1+\bar N(\lambda+\rho X)/\lambda_A$.

In general, the distribution of the \emph{dwell time} $t_D=t_A+t_B$ can be obtained by convolving the probability distributions for $t_A$ and $t_B$. Here we use the exponential distribution Eq.~\reff{exp_A}, and Eq.~\reff{p_t}, respectively, which produce
\be\label{DTD-sol}
\begin{split}
 &p_{\rm D}(t_D)=\int_0^{t_D }dt_B\, p(t_B)p_A(t_D-t_B)\\
&=\lambda_A\left[e^{-\lambda_A t_D}\Big(1+\lambda_A\int_0^{t_D}dt_B\,e^{\lambda_A t_B}P(t_B)\Big)- P(t_D)\right]
\end{split}
\ee
where $P(t)$ is the cumulative distribution of binding times, Eq.~\reff{P(tau)_1codon}.

Note that, as expected, the mean dwell time $\bar t_D$ is simply the sum of the mean binding time $\bar t_B$ and the mean time $\bar t_A$ for the additional biochemical steps, i.e., 
\be\label{mean_dwell_time}
\bar t_D=\frac{\rho^{\rm e}+\lambda^{\rm e}+1}{\bar N (\lambda^{\rm e}+\rho^{\rm e}X)}+\frac{1}{\lambda_A}.
\ee

\section{Discussion and interpretation of the parameters}
\label{parameters}

\begin{table}
\caption{\label{tab:data_ecoli}Data for E.Coli.}
\begin{ruledtabular}
\begin{tabular}{ll}
Typical radius of a ribosome\footnotemark[1] & $L_R\approx 1 \times 10^{-8}\,{\rm m}$\vspace{1mm}\\
\pbox{20cm}{Molar concentration\\of the tRNA\footnotemark[2]} & $c_{\rm t}\approx (0.3\div 30) \times 10^{-6} {\rm M}$\vspace{1mm}\\
\pbox{20cm}{Number concentration\\of the tRNA\footnotemark[2]} & $C_{\rm t}\approx (1.8\div 180) \times 10^{20} {\rm m}^{-3}$\vspace{1mm}\\
\pbox{20cm}{Diffusion constant\\of the tRNA\footnotemark[3]} & $D\approx (0.2\div 2) \times 10^{-12}\,{\rm m}^2{\rm s}^{-1}$\vspace{1mm}\\
Average translation rate\footnotemark[4] & $\Theta\approx 10\div20\, \,{\rm codons}\,{\rm s}^{-1}$\vspace{1mm}\\
Codon length &$\ell_{\rm cod}\approx 1 nm$\vspace{1mm}\\
Total binding rate\footnotemark[5] & $\lambda_B^{\rm tot}\approx xc_{\rm t}\times110\, (\mu {\rm M}\,{\rm s})^{-1}$\vspace{1mm}\\
\end{tabular}
\end{ruledtabular}
\footnotetext[1]{From Ref.~\cite{Zhu1997}}
\footnotetext[2]{Considering a single species of tRNA, from Ref.~\cite{Dong1996}}
\footnotetext[3]{From Ref.~\cite{VanDenBogaart2007}}
\footnotetext[4]{From Ref.~\cite{Bremer1996}}
\footnotetext[5]{From Ref.~\cite{Tinoco2009}; note that $xc_{\rm t}$ is the molar concentration of a charged tRNAs of a specific species.}
\end{table}

In order to understand the physical implications of the findings in the previous sections, it is necessary to obtain an estimate of the parameters $\rho$, $\bar N$, $X$, $\lambda_R$ and $\lambda_B$ in terms of the physical and biological measurable quantities.

Let us first consider the effective volume $V_r$ around the ribosome, as introduced in Sec.~\ref{model}. This volume is delimited by a radius $r$ which corresponds to the maximal distance from the ribosome such that a tRNA has a non-negligible probability of diffusing towards (and being captured by) the ribosome. As shown in Ref.~\cite{Redner2001}, the probability of being absorbed by a target of radius $L_R$ centered at the origin, starting from radius $r'$, is $L_R/r'$. Therefore, we expect that $r$ is on the same order of magnitude of the ribosome radius $L_R$, i.e., $r=\omega L_R$, with $1<\omega < 10$.

The average number of a certain species of tRNAs is found by fixing the concentration in the volume  $V_r=4 \pi r^3/3$ to be the same as in the cell ($C_{\rm t}$). 
We obtain $\bar N= V_r C_{\rm t} \approx \omega^3 (10^{-3}\div 10^{-1})$.
The wide variation is due to two facts: first,  different species of tRNA have very different concentrations. Furthermore, the concentration of a given species of tRNA changes accordingly to the variation in the environmental conditions experienced by the cell.

The parameter $X$ measures the fraction of charged tRNAs in the cell. Its range, measured \emph{in vivo} for E. Coli in Ref.~\cite{Dittmar2005}, spans the interval $X\approx 10^{-3}\div 10^0$, depending on the richness of the growth media. A similar dynamic range was also observed in numerical simulations \cite{Brackley2011}.

The tRNA exchanges between system and reservoir are ruled by $\rho$, which can be obtained in terms of the diffusion constant $D$ of the tRNA molecules, the ribosome velocity $v_{\rm rib}$ and the system size $r$. Supposing that each tRNA performs a Brownian motion, its mean square displacement in the time $T$ is $\ell^2=\langle \sum_i\Delta x_i^2\rangle=6DT$, and the typical exit time from the sphere of radius $r$ is
  \begin{equation}
   T_{\rm{exit}}=\frac{r^2}{6D}.
  \end{equation}
Neglecting the ribosome motion, we equate $T_{\rm exit}$ to the average exit time $1/\rho$ in the stochastic model, obtaining
  \begin{equation}\label{rho_value}
   \rho= \frac{6 D}{r^2}\approx \omega^{-2}(1\div 70) \times 10^3\, {\rm s}^{-1}.
  \end{equation}
The motion of the ribosome produces an additional flux $\rho_T\simeq v_{\rm rib}/r$
\footnote{In the reference frame of the ribosome there is a drift $v_{\rm rib}$ in the tRNAs motion, which determines, in the small time $\delta t$, the exit from the system of an average number $N_{\rm exit}\simeq r^2 v_{\rm rib}\delta t C_{\rm t}\simeq v_{\rm rib}\delta t N/r$ of tRNAs (while, of course, the same average number of tRNAs is entering in the system). The same average number $N_{\rm exit}$ would be caused by a stochastic flux with individual exit rate $\rho_T\simeq v_{\rm rib}/r$}.
Using the data in table \ref{tab:data_ecoli}, the ribosome speed reads $v_{\rm rib}=\Theta\ell_{\rm cod}\approx (10\div 20)\, {\rm nm}/{\rm s}$. 
This produces $\rho_T\approx \omega^{-1}(1\div 2)\, {\rm s}^{-1}$, which is negligible compared to the estimate in Eq.~\reff{rho_value}.

The binding rate $\lambda_B$ can be obtained by equating the total binding rates in our model ($\bar n \lambda_B$) with the experimental one, adapted from Ref.~\cite{Tinoco2009} ($\lambda_B^{\rm tot}$, in Tab.~\ref{tab:data_ecoli}). Solving for $\lambda_B$ produces the estimate $\lambda_B=\lambda_B^{\rm tot}/(xC_{\rm t}V_r)\approx 4\times 10^4 \omega^{-3}\,{\rm s}^{-1}$.

The rate $\lambda_A$ can be readily estimated from the rates of the biochemical steps in Ref.~\cite{Tinoco2009} to be $\lambda_A\approx1.26\,{\rm s}^{-1}$. 
However, this value is not compatible with the average translation rate $\Theta$ measured \emph{in vivo} for E.Coli. In fact, as reported in Tab.~\ref{tab:data_ecoli}, $\Theta\approx (10\div 20)\,{\rm cod/s}$, while from Eq.~\reff{mean_dwell_time}, we expect that $1/\Theta=\bar t\geq1/\lambda_A$.
This lack of consistency between different measures (which also reverberates in a negative estimate of the rescaling factor, that, from Eq.~\reff{gamma}, is $\gamma=\lambda_A/(\lambda_A-\Theta)$)
could be due to the fact that the rates in Ref.~\cite{Tinoco2009} were obtained \emph{in vitro}. 

The last unknown parameter is the recharge rate $\lambda_R$, which can be in principle estimated by restoring in Eq.~\reff{mean_dwell_time} the $\lambda_B$ dependence and equating to the inverse of the average experimental translation rate $\Theta$:
\be\label{Theta_eq}
\frac{1}{\Theta}=\frac{\rho^{\rm e}+\lambda_R^{\rm e}+\lambda_B}{\bar N\lambda_B (\lambda_R^{\rm e}+\rho^{\rm e}X)}+\frac{1}{\lambda_A}.
\ee
The estimate obtained with the values in Tab.~\ref{tab:data_ecoli}, however, is affected by the aforementioned inconsistency.

By measuring  all the involved quantities under the same experimental conditions, it would instead be possible to consistently estimate all the parameters, and therefore to quantify the predicted deviation of the BTD from an exponential distribution.

The deviation of the BTD from the exponential distribution could be also available for a direct experimental measurement with the techniques employed in Refs.~\cite{Uemura2010,Wen2008}:
it would be necessary to introduce the tRNA recharging in the experimental setup (by adding the relative enzymes), and to carefully analyze how and how much the binding time affects the total translation time. The binding time seems to be rate-limiting in the experimental conditions employed in Ref.~\cite{Uemura2010}, as the average translation time changes linearly with the inverse of the tRNA concentration. In this case the BTD should be adequately approximated by Eq.~\reff{p_t}. The same is not true for the experiments in Ref.\cite{Wen2008}, where several timescales are evidently present. 
In this latter case, in order to disentangle the effects due to the binding time from those due to the additional biochemical steps, it would be particularly useful to run a series of experiments at different concentrations of tRNA ($C_{\rm t}$), because a change in $C_{t}$ affects only the binding step, and leaves all the other biochemical steps unchanged.
The interpretation of these results would be potentially possible by refining the framework introduced in Sec.~\ref{other_steps_section}.

\section{Conclusions}
\label{conclusions}

In this paper we develop a microscopic model which describes the binding of the tRNA (charged with the proper amino acid) to the ribosome during the translation of an mRNA sequence into a protein.
This fundamental step heavily depends on the conditions in the cell, and, in particular, on the concentration of charged tRNAs around the ribosome \cite{Uemura2010,Zhang2010,Wohlgemuth2013}.
We consider the recharge dynamics and the diffusion of the tRNA molecules by assuming that each tRNA can be either charged with the relative amino acid, or uncharged. The charging occurs with rate $\lambda_R$, and the charged tRNAs can bind to the ribosome with rate $\lambda_B$. 
Spatial inhomogeneity and stochastic fluctuations of the  number of charged tRNAs around the ribosome are included, and diffusion-driven exchanges  with the reservoir (i.e., the rest of the cell) are allowed. This model neglects the additional biochemical steps which are required to translate a codon, meaning that it is
per se valid in the limit where the binding step is rate limiting. A mean-field approach is introduced in Sec.~\ref{other_steps_section} in order to estimate the effect of these additional reactions.

We describe this non-equilibrium system via its master equation, which in fact violates detailed balance.
We mainly focus on the stationary solution, but we also manage to solve the time dependent master equation from which we extract the relaxation time scales to the stationary solution, and the time correlators for the variables $n$ and $m$ (respectively, the number of charged and uncharged tRNAs in the system).
We are able to obtain the analytical expression of the binding time distribution (BTD), i.e., the distribution of the time intervals spent by the ribosome waiting for a charged tRNA. 
This distribution substantially deviates from the exponential distribution with the same average: specifically, the small and large binding times are over represented in the BTD.
Besides, we numerically checked in a wide range of parameters that the BTD is overdispersed with respect to the exponential distribution with the same mean. 
This fact would be available for experimental measurement with the techniques employed in Refs.~\cite{Uemura2010,Wen2008}, by utilizing experimental conditions such that 
(\emph{i}) the recharge of the tRNAs is allowed and (\emph{ii}) the binding step is rate limiting (as in Ref.~\cite{Uemura2010}).
When the condition (\emph{ii}) is not met, it would still be possible to estimate the effects of the binding time on the total translation time by refining the framework sketched in Sec.~\ref{other_steps_section}; in this case, in order to study the BTD, it would be particularly useful to repeat the experiment with different concentrations of tRNA.

We also show that the appearance of a non exponential BTD is related to the coupling of the fluctuations of $m$ and $n$.
More specifically, we show that the qualitative mechanism is as follows: (\emph{i}) a binding event typically happens when the number of charged tRNAs is risen due to a fluctuation, on average $\bar n_{\textrm{b}^-}=\bar n +1$, (\emph{ii}) during the binding event, a charged tRNA gets discharged and the fluctuation on $n$ propagates to $m$, $\bar m_{\textrm{b}^+}=\bar m +1$, (\emph{iii}) if $\lambda>0$, this fluctuation on $m$ can propagate again on $n$ with a characteristic timescale, producing a "bump" in the timeseries of $n$ as in Fig.~\ref{n_avg_conditioned}.
The size of this effect is larger the smaller the average number of tRNAs $\bar N$ is - i.e., the bigger the relative size of the fluctuations is. 

Concluding, we believe that this kind of models, by analytically dissecting a small set of phenomena, can be very helpful in understanding the quantitative small scale dynamics of the translation process, and in discriminating the main effects from the corrections.

\appendix

\section{Time dependent solution of Eq.~\reff{master_eq_1codon} and two-points correlators}
\label{app:time_dep}

In this appendix we solve the time-dependent master equation, Eq.~\reff{master_eq_1codon}, in order to characterize the relaxation dynamics of the model toward the stationary state. Moreover, by using the properties of the characteristic function, we are able to compute the different-time correlators between $n$ and $m$.

In order to simplify the notation, let us first introduce the quantities
\be
 \begin{split}
  &E=e^{-t(\lambda+\rho+1)},\\
  &F=e^{-\rho t}.
 \end{split} 
\ee
\begin{widetext}
The solution of the differential equation for the generating function $G(z,w;t)$ associated to Eq.~\reff{master_eq_1codon} can be easily obtained with the method of characteristics. Using the initial condition $p_{n,m}(0)=\delta_{n,n_0}\delta_{m,m_0}$, we have
\begin{multline}
\label{G_time_dep}
  G(z,w;t)=\exp\left[ \frac{\bar N\rho(w-z)[X(\lambda+1)-\lambda]}{(\lambda+1)(\lambda+\rho+1)}E+\bar N\frac{1-w+\lambda(1-z)}{\lambda+1}F +\bar N\frac{w-1+\lambda(z-1)+\rho[zX-1+w(1-X)]}{\lambda+\rho+1} \right]\\
\times\left(1+\frac{z-w}{\lambda+1}E+\frac{w-1+\lambda(z-1)}{\lambda+1}F\right)^{n_0}\left(1+\frac{\lambda(w-z)}{\lambda+1}E+\frac{w-1+\lambda(z-1)}{\lambda+1}F\right)^{m_0}.
\end{multline}
By differentiation we obtain the time-dependent probability distribution for $(n,m)$:
\begin{multline}
   p_{n,m}(t)=\left[\frac{(\partial_z)^n}{n!}\frac{(\partial_w)^m}{m!}G(z,w;t)\right]_{z=w=0}=\frac{e^{\bar N(F-1)}}{n!m!(\lambda+1)^{n+m}}\sum_{r=0}^n\sum_{s=0}^mA^{n-r}B^{m-s}
   \sum_{j=0}^r\sum_{k=0}^s\binom{r}{j}\binom{s}{k}\\
\times  \frac{n_0!m_0!\,\lambda^j(E+\lambda F)^{r-j}(\lambda E+F)^k(F-E)^{s+j-k}}{(n_0-r+j-s+k)!(m_0-j-k)!}(1-F)^{n_0+m_0-r-s}\,\theta(n_0-r+j-s+k)\,\theta(m_0-j+k),
\end{multline}
where $\theta(x)$ is the Heaviside step function and
\end{widetext}
\be
 \begin{split}
  &A=(\lambda+1)\bar n-\frac{\bar N\rho[X(\lambda+1)-\lambda]}{\lambda+\rho+1}E-\bar N\lambda F,\\
  &B=(\lambda+1)\bar m+\frac{\bar N\rho[X(\lambda+1)-\lambda]}{\lambda+\rho+1}E-\bar N F.
 \end{split}
\ee

It can be noticed that in the case $n_0=m_0=0$, being the exponent linear in $z$ and $w$, the time-dependent probability distribution is factorized (like the stationary one): $p_{n,m}(t)=p_n(t)p_m(t)$. In this case it reduces to:
\be
 \begin{split}
  &p_{n,m}(t)=\frac{1}{n!}\left[ \bar n-\frac{\bar N}{\lambda+1}\left(\rho\frac{X(\lambda+1)-\lambda}{\lambda+\rho+1}E+\lambda F\right) \right]^n\\
  &\times\frac{1}{m!}\left[ \bar m+\frac{\bar N}{\lambda+1}\left(\rho\frac{X(\lambda+1)-\lambda}{\lambda+\rho+1}E-\lambda F\right) \right]^m e^{-\bar N(1-F)}.
 \end{split}
 \ee

 For generic initial conditions we can write a large-$t$ (i.e., a small $E,F$) expansion, in order to see how $p_{n,m}(t)$ relaxes to the stationary value $p_{n,m}^{\rm st}$, Eq.~\reff{pnm}:
\be\label{pnm_large_t}
 p_{n,m}(t)= p_{n,m}^{\rm st}\left[1+\alpha E+\beta F +O(E^2,F^2,EF)\right],
\ee
where
\be
 \begin{split}
  \alpha=&\frac{\bar N\rho[X(\lambda+1)-\lambda]}{(\lambda+1)(\lambda+\rho+1)}\left(\frac{m}{\bar m}-\frac{n}{\bar n}\right)\\
  &+\frac{n_0\theta(n_0-1)-\lambda m_0\theta(m_0-1)}{\lambda+1}\left(\frac{1}{\bar n}-\frac{1}{\bar m}\right),\\
  \beta=&\bar N-(n_0+m_0)-\frac{\bar N}{\lambda+1}\left(\frac{\lambda n}{\bar n}+\frac{m}{\bar m}\right)\\
  &+\frac{n_0\theta(n_0-1)+m_0\theta(m_0-1)}{\lambda+1}\left(\frac{\lambda}{\bar n}+\frac{1}{\bar m}\right).
 \end{split}
\ee
The leading term for large times is therefore associated with $E$ and $F$, i.e., with the relaxation-times:
\be\label{decay_times}
 \begin{split}
  &t_{1}=\frac{1}{\rho},\\
  &t_{2}=\frac{1}{\lambda+\rho+1}.\\
 \end{split}
\ee
The first time scale $t_1$ is associated with the diffusion process, while the second one is the inverse of the sum of all rates (if we restore the $\lambda_B$ dependence we have $1/t_2=\lambda_R+\rho+\lambda_B$).

Given the analytic expression of the generating function $G(z,w;t)$ in Eq.~\reff{G_time_dep}, it is straightforward to evaluate the correlators, for instance:
 \begin{multline}
\langle n(0)m(t)\rangle=\sum_{n_0,m_0}\sum_{n,m} n_0 m\,p_{n_0,m_0}
  \left[p_{n,m}(t)\right]_{\substack{n(0)=n_0\\m(0)=m_0}}\\
=\sum_{n_0,m_0} n_0 \,p_{n_0,m_0}\big[\partial_wG(z,w;t|n_0,m_0)\big]_{\substack{z=1\\w=1}}.\\
 \end{multline}
In particular, we find:
\be\label{correlators}
 \begin{split}
  &C_{nn}(t)\equiv\langle n(0)n(t)\rangle-\bar n^2=\bar n\frac{\lambda e^{-\rho t}+e^{-t(\lambda+\rho+1)}}{\lambda+1},\\
  &C_{mm}(t)\equiv\langle m(0)m(t)\rangle-\bar m^2=\bar m\frac{e^{-\rho t}+\lambda e^{-t(\lambda+\rho+1)}}{\lambda+1},\\
  &C_{nm}(t)\equiv\langle n(0)m(t)\rangle-\bar n\bar m=\bar n\frac{e^{-\rho t}-e^{-t(\lambda+\rho+1)}}{\lambda+1},\\
  &C_{mn}(t)\equiv\langle m(0)n(t)\rangle-\bar n\bar m=\bar m\lambda\frac{e^{-\rho t}-e^{-t(\lambda+\rho+1)}}{\lambda+1}.\\
 \end{split}
\ee
The first two correlators in \reff{correlators} are monotonically decreasing, as they are linear combinations (with positive coefficients) of the exponentials characterized by the decay times of Eq.~\reff{decay_times}.

The other two correlators are again linear combinations of the same exponentials, but the coefficients of such linear combination have different signs, which makes them non-monotonic. The maximum is at time
\be
\label{tmax}
t_{\rm{max}}=\frac{1}{\lambda+1}\log\left[\frac{\lambda+\rho+1}{\rho}\right].
\ee

\section{Violation of detailed balance}
\label{app:DB}
\begin{figure}
	 \includegraphics[width=0.9\columnwidth]{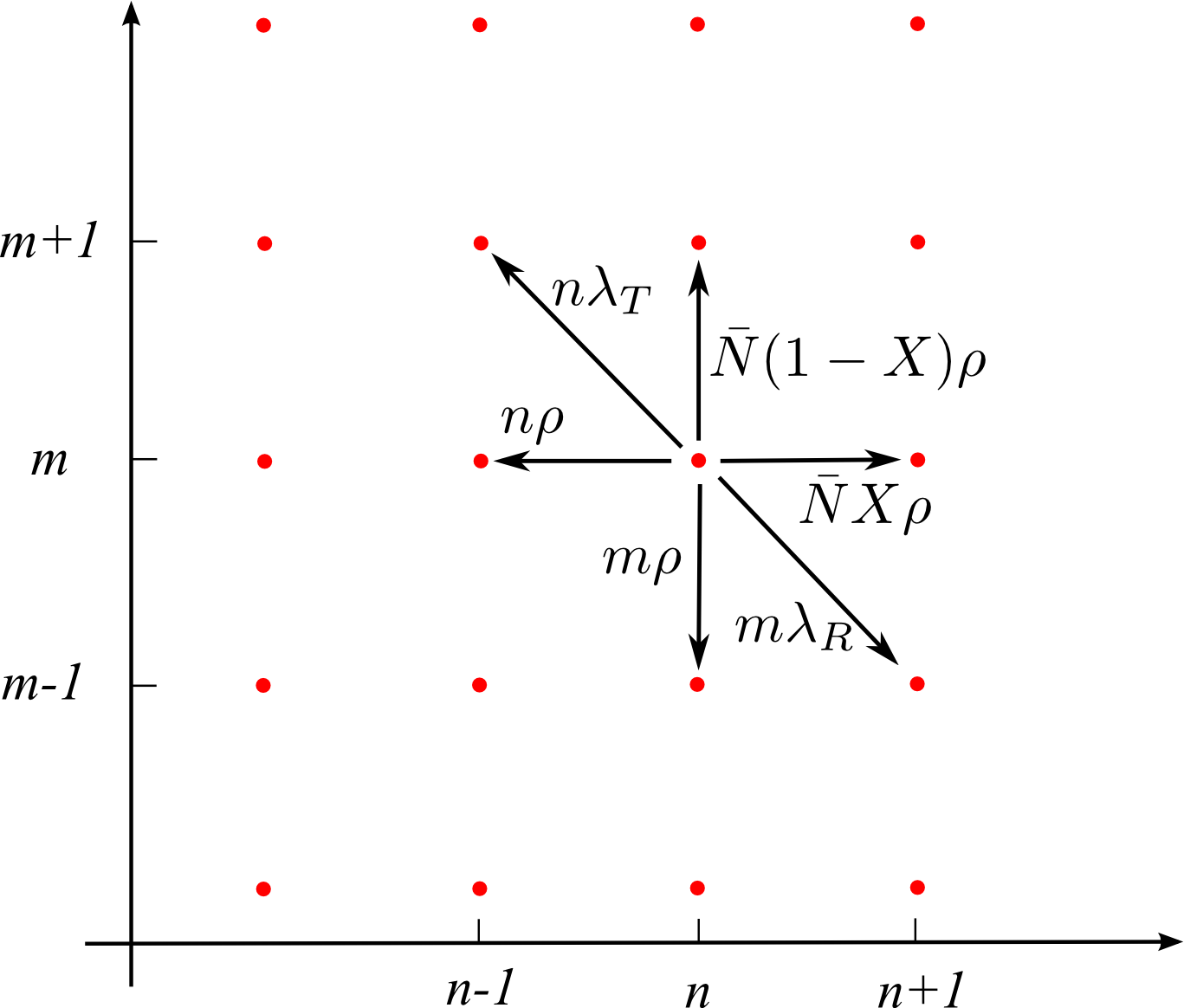}
	 \caption{(Color online) The model as a random walk on the ($n,m$) lattice. The transition rates depend on the direction and on the position on the lattice.}
	 \label{DB:fig}
\end{figure}

The model can be interpreted as a random walk on the  two dimensional $(n,m)$ lattice, with site- and direction-dependent transition rates (see Fig.~\ref{DB:fig}).

We use this analogy to check for eventual violation of detailed balance (DB) in the stationary state described by Eq.~\reff{pnm}. As it can be seen from Fig.~\ref{DB:fig}, there are three directions for the single step jumps:
\begin{itemize}
 \item Along the vertical direction $(n,m)\leftrightarrow(n,m+1)$ the DB condition is
 \be
  (1-X)\rho \bar N\,p_{n,m}=(m+1)\rho\,p_{n,m+1},
 \ee
which is satisfied if
 \be\label{DB_vert}
  \rho=0\hspace{4mm}{\rm or}\hspace{4mm}\rho\to\infty   \hspace{4mm}{\rm or}\hspace{4mm} \lambda=\frac{X}{1-X}.
 \ee
 \item Along the horizontal direction $(n,m)\leftrightarrow(n+1,m)$ the DB condition is
 \be
  X\bar N\rho\,p_{n,m}=(n+1)\rho\,p_{n+1,m}.
 \ee
	Again, its solution is Eq.~\reff{DB_vert}.
 \item Along the diagonal direction $(n,m)\leftrightarrow(n-1,m+1)$ the DB condition is
 \be
  n\,p_{n,m}=\lambda(m+1)\,p_{n-1,m+1},
 \ee
 whose solution is
 \be\label{DB}
 \rho=0\hspace{1cm}{\rm or}\hspace{1cm}\lambda=\frac{X}{1-X}.
 \ee
\end{itemize}
We conclude that the only values of the parameters satisfying DB are those in Eq.~\reff{DB}, while for $\rho\to\infty$ DB is "almost satisfied", being the violation vanishingly small.

The first value of Eq.~\reff{DB} coincides with the trivial case where diffusion is suppressed, while the second one is the value where the stationary points of the internal (recharge and binding) and diffusive dynamics coincide.

In all other cases there are current probability loops and the stationary state is out of equilibrium \cite{Zia2007}.

\section{Relation between the BTD and the average number of charged tRNAs}
\label{app:BTD_n}

Let us consider the time dependent average of the number of charged tRNAs $\langle n(t)\rangle_{\rm nb}=\sum_{n=0}^\infty n \, p_n(t|\textrm{no binding})$, where at $t=0^-$ a binding event occurred and no binding events were recorded between $0$ and $t$. 

In discrete time with temporal step $\delta t$, we can write the probability that a binding event happens at time $t>\tau_k=k\delta t$ as
\begin{equation}
	P(\tau_k)=\prod_{i=0}^k(1-\langle n(\tau_i)\rangle_{\rm nb}\delta t)\sim \exp\left(-\sum_{i=0}^k\langle n(\tau_i)\rangle_{\rm nb}\delta t\right).
\end{equation} 
In the continuous time limit we obtain
\be
\label{eqApp1}
	P(\tau)=\exp\left(-\int_0^\tau dt\, \langle n(t)\rangle_{\rm nb}\right).
\ee
By substituting Eq.~\reff{eqApp1} into \ref{p_t}, we obtain Eq.~\reff{BTD_nt}. Furthermore, by inverting Eq.~\reff{eqApp1}, we can write
\begin{equation}
\label{nAvgCond_P}
	\langle n(t)\rangle_{\rm nb}=-\partial_\tau \log P(\tau)_{|\tau=t}.
\end{equation}
This relation is utilized in Fig.~\ref{n_avg_conditioned} to plot the theoretical predictions.

\begin{acknowledgments}
The authors wish to thank Andrea Gambassi, Giacomo Gori and Matteo Marsili for the useful discussions, and an unknown referee for the valuable comments and suggestions.
This work was partially supported by the GDRE 224 GREFI MEFI, CNRS-INdAM.
\end{acknowledgments}

\bibliography{ref_trnaTransl}

\end{document}